\documentclass[iop]{emulateapj} 
\usepackage{natbib}
\usepackage{graphicx}
\usepackage{amsmath} 
\usepackage{placeins}

\graphicspath{{./}}

\newcommand{\MBH}{M_{\mathrm{BH}}}
\newcommand{\Mhost}{M_{\mathrm{host}}}
\newcommand{\Mbulge}{M_{\mathrm{bulge}}}

\begin{document}

\title{Origin of the Correlations Between Supermassive Black Holes and Their Host Galaxies}

\author
{
Sydney Sherman$^{1,2}$, Mouyuan Sun$^{1,2,3}$, Qirong Zhu$^{1,2}$, Jonathan R. Trump$^{1,2,\dagger}$, Yuexing Li$^{1,2}$ 
}

\affil{$^1$Department of Astronomy \& Astrophysics, The Pennsylvania State University, 
525 Davey Lab, University Park, PA 16802, USA}
\affil{$^2$Institute for Gravitation and the Cosmos, The Pennsylvania State University, University Park, PA 16802, USA}
\affil{$^3$Department of Astronomy and Institute of Theoretical Physics and Astrophysics, Xiamen University, Xiamen, Fujian 361005, China}
\affil{$^\dagger$Hubble Fellow}

\email{sxs5520@psu.edu} 

\begin{abstract}

Observations have shown that supermassive black holes in nearby elliptical galaxies correlate tightly with the stellar velocity dispersion (the $\MBH - \sigma$ relation) and the stellar mass (the $\MBH - \Mhost$ relation) of their host spheroids. However, the origin of these correlations remains ambiguous. In a previous paper by Zhu et al., we proposed a model which links the M-$\sigma$  relation to the the dynamical state of the system and the $\MBH - \Mhost$ relation to the self-regulation of galaxy growth. To test this model, we compile a sample of observed galaxies with different properties and examine the dependence of the above correlations on these parameters. We find that galaxies that satisfy the the $\MBH - \sigma$ correlation appear to have reached virial equilibrium, as indicated by the ratio between kinetic energy and gravitational potential, 2K/U $\sim$ 1. Furthermore, the ratio of black hole accretion rate to star formation rate remains nearly constant, BHAR /SFR $\sim$ $10^{-3}$, in active galaxies over a wide range of mass in the redshift range z=0 - 3. These results confirm our theoretical model that the observed correlations have different origins: the $\MBH - \sigma$ relation may result from galaxy relaxation, while the $\MBH$ - $\Mhost$ relation may be due to self-regulated black hole accretion and star formation in galaxies.

\end{abstract}

\keywords{active galactic nuclei, black holes, black hole -- galaxy scaling relations, galaxy evolution}
 
\section{Introduction}

Over the past decade, it has been determined that a supermassive black hole resides in the center of nearly every galaxy. Furthermore, tight correlations have been found between black hole mass and host galaxy properties (see \citealt{KormendyandHo2013} for a recent review), such as stellar velocity dispersion (the $\MBH - \sigma$ correlation, e.g., \citealt{Ferrarese2000, Gebhardt2000, Tremaine2002, GultekinJune2009, Graham2011}) and bulge stellar mass (the $\MBH - \Mbulge$ correlation, e.g., \citealt{Magorrian1998, Marconi2003, Haring2004}).  Beyond the local universe it is difficult to reliably distinguish the bulge from the rest of the galaxy, and the $\MBH - \Mbulge$ relation is generalized as the $\MBH - \Mhost$ relation between the black hole mass and total stellar mass of the host galaxy. 

Recent studies have placed tighter constraints on the slopes of the relationships by obtaining more accurate black hole and galaxy measurements to understand the evolution of the M - $\sigma$ and $\MBH$ - $\Mhost$ relations (e.g., \citealt{Peng2006, Treu2007, GultekinJune2009, Jahnke2009, Merloni2010, Bennert2011, SchrammSilverman2013, McConnellMa2013, Sun2014}). Observational studies beyond the local universe are complicated by a significant Eddington bias caused by the steeply declining black hole and galaxy mass functions coupled with large errors in estimating $\MBH$ and intrinsic scatter in $\MBH$ - $\Mhost$ (\citealt{Lauer2007, ShenKelly2010}).  In addition, for galaxies beyond $\sim$200~Mpc $\MBH$ can only be estimated in luminous broad-line AGNs, which frequently outshine their host galaxies and make it difficult to measure half light radius and stellar velocity dispersion for the host galaxy.  Most observations which account for these challenges find little evidence for evolution in $\MBH$ - $\Mhost$ relation from $z \sim 2$ to $z \sim 0$ (e.g. \citealt{Jahnke2009, SchrammSilverman2013, Sun2014}).

Multi-wavelength surveys in deep \textit{Hubble Space Telescope} (\textit{HST}) fields allow for detailed studies of AGN and host galaxy properties. More specifically, the black hole accretion rate (BHAR) can be reliably determined using X-rays, and far-infrared (FIR) data allow for the calculation of a host galaxy's star formation rate (SFR) with no contamination from the AGN (e.g., \citealt{Mullaney2012b, Rosario2013a, Rosario2013b, Sun2014}). Both of these properties are useful in understanding the evolution of a galaxy and its supermassive black hole. With further constraints on the BHAR and SFR, it can be determined if evolution --- or lack therof --- is actually coevolution.  Recent observations suggest that there is a correlation between BHAR and SFR in an \textit{average} sense (\citealt{Mullaney2012b, Chen2013, Sun2014}), but with less evidence for a connection in \textit{individual} galaxies (\citealt{Mullaney2012a}), possibly due to AGN variability (\citealt{Hickox2014}). In particular, \cite{Sun2014} combined robust measurements of both $\MBH$ - $\Mhost$ and BHAR - SFR, along with careful modeling of the selection biases, to demonstrate that black hole accretion and star formation correlate in such a way to ``self-maintain'' the observed constant $\MBH$ - $\Mhost$ relation.

Despite advancements in measurement techniques, the physical origins of these correlations remains unclear (for a review, see Section 8 of \citealt{KormendyandHo2013}). Recently, \cite{Zhu2012} studied the BH -- galaxy relations with high-resolution cosmological simulations and found that the $\MBH - \sigma$  relation evolves significantly with redshift, but the $\MBH$ - $\Mhost$ relation shows little evolution. The authors suggested that the $\MBH - \sigma$ and $\MBH - \Mhost$ relations have different origins, the former being a result of virial equilibrium of the system, while the latter owing to self-regulated growth of galaxies.

In this work, we focus on determining the origins of both the $\MBH$ - $\sigma$ and the $\MBH$ - $\Mhost$ relations and testing the model of \cite{Zhu2012} by compiling a sample of galaxies with different properties (e.g., mass, type, kinematics, etc.) and at different redshift, and examine the dependence of the above correlations on these parameters. In Section 2, we outline the samples used in our study. The results are presented in Section 3, and we summarize in Section 4. 

\section{Samples}
\label{sec:samples}

In order to effectively study the origins of the $\MBH$ - $\sigma$ and the $\MBH$ - $\Mhost$ relations, two separate samples of galaxies were required. The first sample, used to study the $\MBH$ - $\sigma$ correlation consists of elliptical, lenticular, and spiral galaxies (tables 1 and 2). The second sample contains spiral galaxies and AGN, and is used to study the $\MBH$ - $\Mhost$ correlation (Table 3, \citealt{Lusso2011} and \citealt{Sun2014} samples). 

Sample 1 is drawn from five recent studies (\citealt{Marconi2003, GultekinJune2009, Kormendy2011, Greene2010, Sani2011}) of the $\MBH$ - $\sigma$ and $\MBH$ - $\Mhost$ correlations, and contains 72 galaxies. These papers were chosen because they presented galaxy type, distance, black hole mass, k-band absolute magnitudes, and stellar velocity dispersions for each of the galaxies. The final parameter of interest to this study, the half-light effective radius, was not presented in all five of the papers, and was therefore obtained from other sources. The galaxy parameters and their references are detailed in Tables 1 and 2. 

In order to study the $\MBH$ - $\Mhost$ correlation, it was necessary to have X-ray and FIR luminosities, which are tracers of BHAR and SFR (\citealt{Kennicutt1998}) respectively. While we were able to obtain X-ray luminosities for the majority of galaxies in our elliptical sample, it is likely that the X-ray luminosity was produced by hot gas rather than BH accretion. Similarly, FIR (as well as UV, H$\alpha$ and other SFR tracers \citealt{Kennicutt1998}) luminosities for these ellipticals were not likely to be due to star formation, but other processes. In order to overcome this BHAR and SFR tracer problem, we created a separate sample that contains active galaxies with robust SFR and BHAR estimates. 

Sample 2 contains 70 broad-lined AGN from \cite{Sun2014}, 11 narrow-lined AGN from \cite{Hainline2012}, 17 narrow-lined AGN from \cite{Lusso2011}, and 5 spiral galaxies from Sample 1 that had published SFR and X-ray luminosities. In this sample, the only galaxies with BH masses are those from \cite{Sun2014} sample and the five spiral galaxies. All other galaxies in Sample 2 are narrow-lined, and do not have BH mass estimates. Regardless of the original sample from which the galaxies were taken, all of the galaxies in Sample 2 have SFR values, X-ray (or bolometric) luminosity, and galaxy stellar mass. The \cite{Hainline2012} values come directly from Table 4 of that paper, while our \cite{Lusso2011} sample was taken from a larger sample of 225 galaxies from the \cite{Lusso2011} paper. We obtained the 17 galaxies from \cite{Lusso2011} by first removing any galaxies that were missing measurements for bolometric luminosity, SFR, stellar mass, or redshift. We then removed galaxies in all but three morphological classes (disk-dominated, irregular, and compact/irregular) that were defined by \cite{Lusso2011}. Finally, the five spiral galaxies were obtained from our larger collection of spirals in Sample 1 because they had both SFR and X-ray luminosity measurements.

\begin{deluxetable*}{cccccccc}[h!]
\tablecaption{\label{tab:sims}Elliptical/Lenticular Sample}
\tablehead{
Galaxy Name\tablenotemark{[1]}&
Type\tablenotemark{[2]}&
D\tablenotemark{[3]}&
$M_{BH}$\tablenotemark{[4]}&
$\sigma$\tablenotemark{[5]} &
$M_K$\tablenotemark{[6]}&
$r_e$\tablenotemark{[7]}&
Ref{[8]}\\
				& 		&(Mpc) 	&($M_{\odot}$)		&(km $s^{-1}$) 	&		&(kpc) &		\\
}
\hline
A1836-BCG		&E		&157.5	&3.9 x $10^9$		&288		&-25.95			&0.099	&2,10,5	\\
A3565-BCG		&E		&54.4	&5.2 x $10^8$ 		&322		&-25.98			&0.0916	&2,10,5	\\
CenA			&S0/E	&4.4		&3.0 x $10^8$		&150		&-22.94			&2.21	&3		\\
Cygnus A			&E		&240		&2.9 x $10^9$		&270		&-27.3			&31.0	&1		\\
IC 1459			&E4		&30.9	&2.8 x $10^9$		&340		&-25.69			&9.15	&3		\\
IC 4296			&E		&50.8	&1.35 x $10^9$		&226		&-27.62			&8.28	&3		\\
M84/NGC 4374		&E1		&18.4	&1.0 x $10^9$		&296		&-25.7			&8.2		&1		\\
M87				&E1		&17.0	&3.6 x $10^9$		&375		&-25.37			&8.2		&3		\\
NGC 1399		&E1		&21.1	&5.1 x $10^8$		&337		&-25.29			&3.63	&2,10,6	\\
NGC 221			&E2		&0.86	&3.1 x $10^6$		&75		&-18.79			&0.12	&3		\\
NGC 2549		&S0		&12.3	&1.4 x $10^7$		&145		&-22.17			&0.69	&3		\\
NGC 2778		&E2		&22.9	&1.4 x $10^7$		&175		&-23.0			&3.0		&1		\\
NGC 2974		&E4		&21.5	&1.7 x $10^8$		&227		&-24.09			&2.83	&3		\\
NGC 3115			&S0		&9.7		&9.1 x $10^8$		&230		&-24.4			&4.7		&1		\\
NGC 3245		&S0		&20.9	&2.1 x $10^8$		&205		&-23.75			&1.3		&4,4,1	\\
NGC 3377		&E5		&11.2	&1.0 x $10^8$		&145		&-23.6			&5.4		&1		\\
NGC 3379		&E0		&11.7	&1.2 x $10^8$		&206		&-24.2			&1.7		&2,1,6	\\
NGC 3384		&S0		&11.6	&1.6 x $10^7$		&143		&-22.6			&0.49	&1		\\
NGC 3414		&S0		&25.2	&2.51 x $10^8$		&205		&-23.49			&2.67	&3		\\
NGC 3585		&S0		&21.2	&3.4 x $10^8$		&213		&-24.67			&1.59	&3		\\
NGC 3607		&E1		&19.9	&1.2 x $10^8$		&229		&-24.44			&4.3		&3		\\
NGC 3608		&E1		&23.0	&2.1 x $10^8$		&182		&-23.74			&6.29	&3		\\
NGC 3998		&S0		&14.9	&2.4 x $10^8$		&305		&-23.29			&0.34	&3		\\
NGC 4026		&S0		&15.6	&2.1 x $10^8$		&180		&-23.14			&0.86	&3		\\
NGC 4261		&E2		&33.4	&5.5 x $10^8$		&315		&-25.27			&3.66	&3		\\
NGC 4291		&E2		&26.2	&3.1 x $10^8$		&242		&-23.9			&2.3		&1		\\
NGC 4342		&S0		&18.0	&3.6 x $10^8$		&225		&-22.26			&0.29	&4,4,1	\\
NGC 4459		&E2		&17.0	&7.4 x $10^7$		&167		&-24.5			&3.24	&2,1,7	\\
NGC 4473		&E4		&17.0	&1.3 x $10^8$		&190		&-23.8			&3.98	&2,1,8	\\
NGC 4486A		&E2		&17.0	&1.3 x $10^7$		&111		&-21.8			&0.65	&2,3,9	\\
NGC 4552		&E		&15.3	&5.0 x $10^8$		&252		&-24.32			&1.8		&3		\\
NGC 4564		&S0		&17.0	&6.9 X $10^7$		&162		&-23.4			&1.59	&2,1,8	\\
NGC 4621		&E5		&18.3	&4.0 x $10^8$		&225		&-23.64			&5.46	&3		\\
NGC 4649		&E2		&16.5	&2.1 x $10^9$		&385		&-25.37			&3.77	&3		\\
NGC 4697		&E6		&12.4	&2.0 x $10^8$		&177		&-23.02			&6.04	&3		\\
NGC 4742		&E4		&15.5	&1.4 x $10^7$		&90		&-23.0			&2.0		&1		\\
NGC 5077		&E3		&44.9	&8.0 x $10^8$		&222		&-25.09			&6.35	&3		\\
NGC 524			&S0		&33.6	&8.32 x $10^8$		&235		&-25.37			&4.37	&3		\\
NGC 5252		&S0		&96.8	&1.0 x $10^9$		&190		&-25.6			&9.7		&1		\\
NGC 5576		&E3		&27.1	&1.8 x $10^8$		&183		&-22.97			&4.51	&3		\\
NGC 5813		&E1		&32.2	&7.1 x $10^8$		&230		&-24.37			&17.4	&3		\\
NGC 5845		&E3		&28.7	&2.9 x $10^8$		&234		&-23.27			&0.51	&3		\\
NGC 5846		&E0		&24.9	&1.1 x $10^9$		&237		&-25.04			&4.4		&3		\\
NGC 6251		&E1		&106		&6.0 x $10^8$		&290		&-26.24			&21.8	&3		\\
NGC 7052		&E3		&70.9	&4.0 x $10^8$		&266		&-25.19			&13.5	&3		\\
NGC 7457		&S0		&14.0	&4.1 x $10^6$		&67		&-21.8			&4.31	&2,1,8	\\
NGC 821			&E4		&25.5	&4.2 x $10^7$		&209		&-23.720			&7.86	&3		\\				
\hline
\tablenotetext{[1]}{Galaxy Name.}
\tablenotetext{[2]}{Morphological type.} 
\tablenotetext{[3]}{Distance in Megaparsecs.}
\tablenotetext{[4]}{BH mass in units of solar mass.}
\tablenotetext{[5]} {Stellar velocity dispersion.}
\tablenotetext{[6]}{K Band bulge magnitude.}
\tablenotetext{[7]}{Effective radius in kiloparsecs.}
\tablenotetext{[8]}{References are written as: ``reference for columns 1-5,"``reference for column 6," ``reference for column 7". If there is only one number, it indicates  the reference for columns 1-7. References: 1. \cite{Marconi2003}, 2. \cite{GultekinJune2009}, 3. \cite{Sani2011}, 4. \cite{Kormendy2011}, 5.\cite{DallaBonta2009}, 6. \cite{Faber1997}, 7. \cite{Kormendy2010}, 8.\cite{Novak2006}, 9. \cite{Prugniel2011}, 10. \cite{KormendyandHo2013}.}
\end{deluxetable*}

\begin{deluxetable*}{cccccccccc}[h!]
\tablecaption{\label{tab:sims}Spiral Sample}
\tablehead{
Galaxy Name\tablenotemark{[1]}&
Type\tablenotemark{[2]}&
D\tablenotemark{[3]}&
$M_{BH}$\tablenotemark{[4]}&
$\sigma$\tablenotemark{[5]}& 
$M_K$\tablenotemark{[6]}&
$r_e$\tablenotemark{[7]}&
Ref{[8]}&
$L_x$\tablenotemark{[9]}&
SFR\tablenotemark{[10]}
\\
				& 			&(Mpc) &($M_{\odot}$)		&(km $s^{-1}$) &(kpc) &	& 		&(ergs $s^{-1}$)	&($M_{\odot}$ $yr^{-1}$)	\\
}
\hline
Circinus			&Sb			&4.0		&1.7 x $10^6$		&158		&-21.79	&0.21	&3 		&41.48 			&1.28\\
IC 2560			&SBb		&40.7	&4.4 x $10^6$		&144		&-22.87	&5.43	&3		&-				&-	\\
M31/NGC 0224		&Sb			&0.8		&4.5 x $10^7$		&160		&-22.8	&1.0		&1		&-				&-	\\
Milky Way			&SBbc		&0.008	&4.1 x $10^6$		&103		&-22.3	&0.7		&1		&-				&-	\\
NGC 1023		&SB0		&12.1	&4.0 x $10^7$		&205		&-23.07	&1.41	&3		&-				&-	\\
NGC 1068		&Sb			&15.4	&8.6 x $10^6$		&151		&-23.79	&0.73	&3		&39.54 			&0.54\\
NGC 1300		&SB(rs)bc		&20.1	&7.1 x $10^7$		&218		&-21.71	&0.422	&2,15,7	&39.93			&0.2	\\
NGC 1316		&SB0		&19.0	&1.62 x $10^8$		&226		&-24.72	&8.57	&3		&-				&-	\\
NGC 2273		&SB(r)a:		&26.0	&7.58 x $10^6$		&144.54	&-22.07	&1.97	&4,15,9	&-				&-	\\
NGC 2748		&Sc			&24.9	&4.7 x $10^7$		&115		&-20.56	&0.82	&2,15,14	&-				&-	\\
NGC 2787		&SB0		&7.9		&4.3 x $10^7$		&189		&-21.37	&0.6		&3		&-				&-	\\
NGC 2960		&Sa?		&71.0	&1.12 x $10^7$		&165.96	&-24.36	&0.62	&4,15,10	&-				&-	\\
NGC 3031		&Sb			&4.1		&8.0 x $10^7$		&143		&-22.82	&2.53	&3		&-				&-	\\
NGC 3079		&SBcd		&15.9	&2.5 x $10^6$		&146		&-22.37	&5.71	&3		&-				&-	\\
NGC 3227		&SBa		&17.0	&2.0 x $10^7$		&133		&-21.52	&6.83	&3		&41.55 			&0.48\\
NGC 3384		&SB0		&11.7	&1.0 x $10^7$		&143		&-22.82	&0.25	&3		&-				&-	\\
NGC 3393		&(R')SB(rs)	&53.6	&3.09 x $10^7$		&147.91	&-23.03	&2.26	&4,15,11	&-				&-	\\
NGC 4151		&Sa			&20.0	&6.5 x $10^7$		&156		&-22.39	&0.52	&3		&-				&-	\\
NGC 4258		&SABbc		&7.2		&3.78 x $10^7$		&115		&-21.54	&3.9		&3		&-				&-	\\
NGC 4388		&SA(s)b:		&19.0	&8.51 x $10^6$		&107.15	&-20.55	&3.33	&4,15,12	&-				&-	\\
NGC 4594		&Sa			&9.8		&1.0 x $10^9$		&240		&-25.4	&5.1		&1		&-				&-	\\
NGC 4596		&SB0		&18.0	&8.4 x $10^7$		&136		&-22.82	&2.44	&3		&-				&-	\\
NGC 6264		&S?			&136	.0	&2.81 x $10^7$		&158.49	&-22.6	&6.43	&4,15,13	&-				&-	\\
NGC 7582		&SBab		&22.3	&5.5 x $10^7$		&156		&-21.96	&0.712	&2,15,8	&41.69 			&2.22\\		
\hline
\tablenotetext{[1]}{Galaxy Name.}
\tablenotetext{[2]}{Morphological type.} 
\tablenotetext{[3]}{Distance in Megaparsecs.}
\tablenotetext{[4]}{BH mass in units of solar mass.}
\tablenotetext{[5]} {Stellar velocity dispersion.}
\tablenotetext{[6]}{K Band bulge magnitude.}
\tablenotetext{[7]}{Bulge effective radius in kiloparsecs.}
\tablenotetext{[8]}{References are written as: ``reference for columns 1-5,"``reference for column 6," ``reference for column 7". If there is only one number, it indicates  the reference for columns 1-7. References: 1. \cite{Marconi2003}, 2. \cite{GultekinJune2009}, 3. \cite{Sani2011}, 4. \cite{Greene2010}, 6.\cite{Erwin2003}, 7. \cite{Novak2006}, 8. \cite{Faber1997}, 9. \cite{Dong2006}, 10. \cite{Vika2012}, 11. \cite{Beifiori2009}, 12. \cite{Khorunzhev2012}, 13. \cite{PerezGonzales2000} , 14. \cite{GildePaz2006}, 15. \cite{KormendyandHo2013}.}
\tablenotetext{[9]}{X-ray luminosity . All are from \cite{GultekinNovember2009} excluding NGC 1300, which is from \cite{Shapley2001}. }
\tablenotetext{[10]}{Star Formation Rate. Values for Circinus, NGC 1068, NGC 3227, and NGC 7582 are from \cite{Melendez2008}. The value for NGC 1300 is from \cite{Mazzuca2008}.}

\end{deluxetable*}

\section{Results}
\label{sec:results}

\subsection{The M-$\sigma$ Correlation}

Figure 1 shows the well-studied M-$\sigma$ correlation from our compiled sample. The red and blue fits correspond to the spirals and ellipticals respectively, while the black dashed-line is from \cite{McConnellMa2013}.  It can clearly be seen that the elliptical galaxies more closely follow the fit from \cite{McConnellMa2013} than the spiral sample. The fitting line for the spirals noticeably diverges from the \cite{McConnellMa2013} fit and has a much shallower slope. The differing fit lines for spirals and ellipticals suggests that the elliptical galaxies more closely follow the M-$\sigma$ correlation than spiral galaxies. 

While there is some deviation from the trend lines in both the elliptical and spiral samples, this can be explained by the inconsistencies in the methods used to determine BH mass and $\sigma$ values across the multiple samples from which these values were obtained. The most common of which are dynamical measurements (\citealt{Marconi2003, Kormendy2011, Sani2011, GultekinJune2009}) and megamaser disk measurements (e.g., \citealt{Greene2010}).

\begin{figure}[h]
\begin{center}
\includegraphics[width=3.3in]{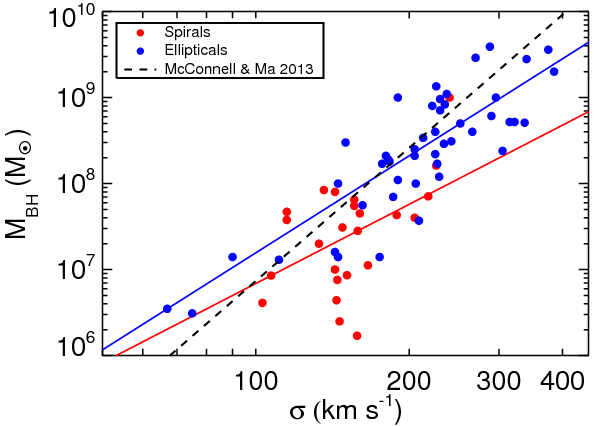} 
\caption{The $\MBH$-$\sigma$ correlation for the spiral and elliptical samples. The red and blue fits correspond to the spirals and ellipticals respectively. It can be seen that the ellipticals more closely follow the fit of \cite{McConnellMa2013} than the spiral galaxies.}
\label{fig_acc_t}
\end{center}
\end{figure}

In \cite{Zhu2012}, it was suggested that M-$\sigma$ correlation is a result of virial equilibrium when 2K + U= 0, where $$K = \frac{3}{2}\text{$M_*$}\sigma^2$$ is the kinetic energy of the stars, and $$U = \frac{-3}{5}\frac{\text{GM$M_*$}}{\text{$r_e$}}$$ is the gravitational potential energy of the galaxy of mass M with effective radius $r_e$.  To test this, we examine the virial ratio, $\lambda =2K/U$, of our sample in Figure~2. For simplicity, we assumed that the mass of the galaxy (M) was equal to the stellar mass ($M_*$), and therefore, our final calculation used $$U = \frac{-3}{5}\frac{\text{G$M_*^2$}}{\text{$r_e$}}$$ for the potential energy. The stellar mass for this figure, and all subsequent figures, is calculated from k-band absolute magnitude using a one-to-one mass to light ratio. 

The results of \cite{Zhu2012} suggest that the M-$\sigma$ relation evolves over time, and subsequently, the virial ratio of galaxies evolves with the M-$\sigma$ relation. Ultimately, if galaxies become virilized as they approach z $\sim$ 0, the virial ratio would reach $\sim$1. Considering that we find that ellipticals more closely follow the M-$\sigma$ correlation, we would predict that they have a virial ratio of $\sim$1, and therefore, we would expect spiral galaxies to have a virial ratio greater than 1. In our plot, however, we see that the elliptical and spiral samples seem to have comparable scatter and virial ratios. While this seems odd because we would expect very different ratios, the story is not complete. Because spiral galaxies do not have uniform light curves, it is difficult to define a ``half-light" radius for them. The majority of effective radius values that were obtained for our spiral sample are half light radii for the bulge, and these are the galaxies with virial ratios of $\sim$1. The spiral galaxies with ``effective radii" for the entire galaxy have a virial ratio much higher than 1. The scatter seen in the elliptical sample is likely due to the difficult nature of measuring the effective radius and our generalization that a one-to-one mass-to-light ratio could be used for the calculation of all galaxy masses.  

\begin{figure}[h]
\begin{center}
\includegraphics[width=3.3in]{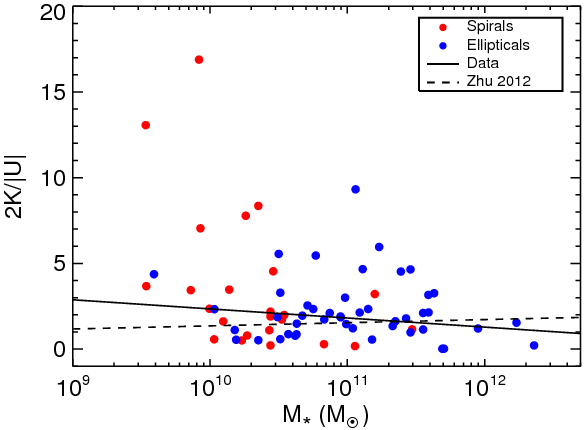} 
\caption{The ratio of kinetic energy to potential energy of stellar components within the half-light radius as a function of stellar mass for our elliptical and spiral samples. Those values with a ratio of roughly 1 are described as virilized, while those that lie significantly above 1 are not virilized.}  
\label{fig_acc_tn}
\end{center}
\end{figure}

\subsection{The $M_{BH}$-$M_{host}$ Correlation}
Figure 3 utilizes the BH masses published by \cite{Sun2014} and those for the five spirals with SFR and X-ray luminosity, as well as the stellar masses for these galaxies to study the $\MBH$ - $\Mhost$ relation. 

\begin{figure}[h]
\begin{center}
\includegraphics[width=3.3in]{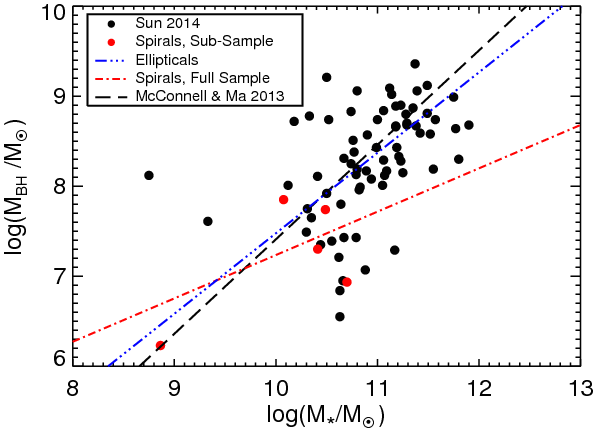}
\caption{BH mass as a function of stellar mass for \cite{Sun2014} and the five spiral galaxies with SFR and X-ray luminosities. The fittings of the full elliptical and spiral samples (Tables 1 and 2) are plotted for comparison.} 
\label{fig_tri_n}
\end{center}
\end{figure}

\begin{figure*}[t!]
\begin{center}
\includegraphics[width=3.3in]{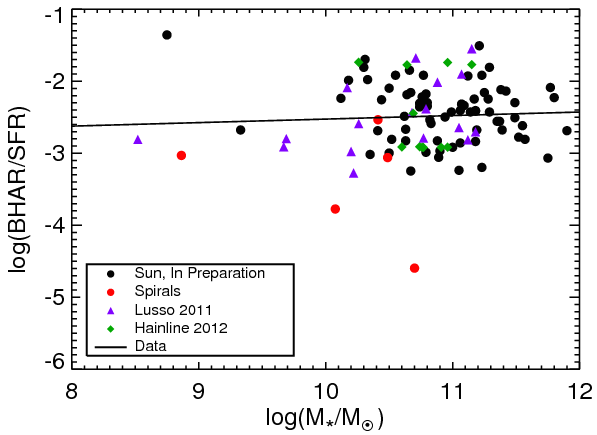}
\includegraphics[width=3.3in]{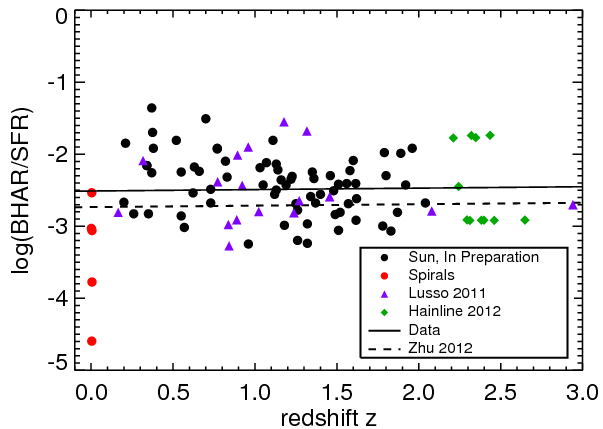}
\caption{Left:  The ratio of BHAR/SFR as a function of stellar mass. Our nearly constant ratio of $\sim$$10^{-2.6}$ across four orders of magnitude in stellar mass suggests that galaxies and black holes have self-regulated growth. Right: The same ratio of BHAR/SFR as a function of redshift. This expands on the work of \cite{Silverman2009}, and provides evidence for self regulated growth of black holes and galaxies to redshift 3.}     
\label{fig_acc_n}
\end{center}
\end{figure*}

While \cite{Zhu2012} found evolution in the M - $\sigma$ correlation, no evolution in the $M_{BH}$-$M_{host}$ correlation was found, which implies that the two correlations have very different origins. The fourth and final figure probes the possible origin of the $M_{BH}$-$M_{host}$ correlation by utilizing the ratio of BH accretion rate (BHAR) to star formation rate (SFR) as a function of stellar mass and redshift for Sample 2. This ratio allows us to track the evolution of the galaxies. 

SFR values for each of our AGN were taken from \cite{Sun2014}, \cite{Lusso2011}, and \cite{Hainline2012}. The references for each of our five spirals are given in detail in Table 2. The BHAR was calculated using $$ \text{BHAR} = \frac{\text{$L_{bol}$}}{c^2\epsilon} $$ where $\epsilon$ = 0.1. X-ray luminosity was converted to bolometric luminosity using $$\text{$L_{bol}$}=\text{$L_x$}*22.4$$ from \cite{Mullaney2012} when direct $L_{bol}$ measurements were unavailable. 

\cite{Zhu2012} determined through simulation that galaxies have a nearly constant ratio of BHAR to SFR over a large range of redshifts, suggesting that black holes and galaxies exhibit self-regulated growth. \cite{Silverman2009} came to similar conclusions studying the ratio of BHAR to SFR using observational data to z=2. Our sample extends the study of \cite{Silverman2009} to z=3, and further constrains the value of this ratio to $\sim$$10^{-2.6}$. This constant ratio across a range of redshifts and stellar masses suggests that the galaxies and black holes exhibit self regulated growth.

\section{Summary}

Through this observational data, we investigated the origin of the $M_{BH}$-$\sigma$ and $M_{BH}$-$M_{host}$ correlations and tested the model proposed by \cite{Zhu2012}. To study the $\MBH$ - $\sigma$ correlation, we have used robust sample of elliptical, lenticular, and spiral galaxies. We have shown that the elliptical and lenticular galaxies closely follow the most recent fitting of the $\MBH$ - $\sigma$ correlation from \cite{McConnellMa2013}, while the spiral galaxies tend to fall off of this correlation. In order to examine the $M_{BH}$ - $M_{host}$ relation, we have used a sample that combines narrow- and broad-lined AGN with a small number of spiral galaxies. We have shown that the broad-lined AGN and spirals follow the $M_{BH}$ - $M_{host}$ from \cite{McConnellMa2013}. Our findings are listed below:

\begin{enumerate}

\item \cite{Zhu2012} has shown that the M-$\sigma$ relation shows exceptional evolution from higher redshift to lower redshift, which suggests that the virial ratio would also evolve over time. Through simulation, \cite{Zhu2012} determined that galaxies progress towards a virial ratio of $\sim$1 at z$\sim$0. Our observational data agrees with the results of \cite{Zhu2012}, with some scatter caused by measurement uncertainties. At z$\sim$0, we see that galaxies falling on the $\MBH$ - $\sigma$ relation tend to be viralized, while those that fall off of the $\MBH$ - $\sigma$ relation are not viralized. This suggests that the $\MBH$ - $\sigma$ relation may be a result of virial equilibrium. 

\item In contrast to the distinct evolution of the M-$\sigma$ correlation, \cite{Zhu2012} found almost no evolution in the $M_{BH}$-$M_{host}$ correlation. We utilize the ratio of BHAR/SFR as a probe of galaxy evolution, and find that the ratio remains relatively constant over several orders of magnitude in stellar mass, and a range of redshifts (z=0-3). Our nearly constant ratio of $\sim$$10^{-2.6}$ agrees with the findings of \cite{Zhu2012} and \cite{Silverman2009}. This nearly constant ratio indicates that the black hole and galaxy have self-regulated growth, and it is this self-regulated growth that may be the origin of the $M_{BH}$-$M_{host}$ correlation.

\end{enumerate}

\section{Acknowledgments}
We thank Jenny Greene, Louis Ho, and Don Schneider for useful discussions. SS acknowledges support from the Penn State University Office of Undergraduate Education. MYS acknowledges support from the China Scholarship Council (No. [2013]3009), and the National Natural Science Foundation of China under grant \#11222328. JRT acknowledges support from NASA through Hubble Fellowship grant \#51330 awarded by the Space Telescope Science Institute, which is operated by the Association of Universities for Research in Astronomy, Inc., for NASA under contract NAS 5-26555. YL acknowledges partial support from NSF grants AST-0965694, AST-1009867 and AST-1412719. We acknowledge the Institute For CyberScience at The Pennsylvania State University for providing computational resources and services that have contributed to the research results reported in this paper. The Institute for Gravitation and the Cosmos is supported by the Eberly College of Science and the Office of the Senior Vice President for Research at the Pennsylvania State University.


\nocite{*}
\begin{thebibliography}{77}
\expandafter\ifx\csname natexlab\endcsname\relax\def\natexlab#1{#1}\fi

\bibitem[{{Baggett} {et~al.}(1998){Baggett}, {Baggett}, \&
  {Anderson}}]{Baggett1998}
{Baggett}, W.~E., {Baggett}, S.~M., \& {Anderson}, K.~S.~J. 1998, \aj, 116,
  1626

\bibitem[{{Beifiori} {et~al.}(2009){Beifiori}, {Sarzi}, {Corsini}, {Dalla
  Bont{\`a}}, {Pizzella}, {Coccato}, \& {Bertola}}]{Beifiori2009}
{Beifiori}, A., {Sarzi}, M., {Corsini}, E.~M., {Dalla Bont{\`a}}, E.,
  {Pizzella}, A., {Coccato}, L., \& {Bertola}, F. 2009, \apj, 692, 856

\bibitem[{{Bennert} {et~al.}(2009){Bennert}, {Barvainis}, {Henkel}, \&
  {Antonucci}}]{Bennert2009}
{Bennert}, N., {Barvainis}, R., {Henkel}, C., \& {Antonucci}, R. 2009, \apj,
  695, 276

\bibitem[{{Bennert} {et~al.}(2011){Bennert}, {Auger}, {Treu}, {Woo}, \&
  {Malkan}}]{Bennert2011}
{Bennert}, V.~N., {Auger}, M.~W., {Treu}, T., {Woo}, J.-H., \& {Malkan}, M.~A.
  2011, \apj, 742, 107

\bibitem[{{Bongiorno} {et~al.}(2012){Bongiorno}, {Merloni}, {Brusa},
  {Magnelli}, {Salvato}, {Mignoli}, {Zamorani}, {Fiore}, {Rosario}, {Mainieri},
  {Hao}, {Comastri}, {Vignali}, {Balestra}, {Bardelli}, {Berta}, {Civano},
  {Kampczyk}, {Le Floc'h}, {Lusso}, {Lutz}, {Pozzetti}, {Pozzi}, {Riguccini},
  {Shankar}, \& {Silverman}}]{Bongiorno2012}
{Bongiorno}, A., {Merloni}, A., {Brusa}, M., {Magnelli}, B., {Salvato}, M.,
  {Mignoli}, M., {Zamorani}, G., {Fiore}, F., {Rosario}, D., {Mainieri}, V.,
  {Hao}, H., {Comastri}, A., {Vignali}, C., {Balestra}, I., {Bardelli}, S.,
  {Berta}, S., {Civano}, F., {Kampczyk}, P., {Le Floc'h}, E., {Lusso}, E.,
  {Lutz}, D., {Pozzetti}, L., {Pozzi}, F., {Riguccini}, L., {Shankar}, F., \&
  {Silverman}, J. 2012, \mnras, 427, 3103

\bibitem[{{Caccianiga} {et~al.}(2007){Caccianiga}, {Severgnini}, {Della Ceca},
  {Maccacaro}, {Carrera}, \& {Page}}]{Caccianiga2007}
{Caccianiga}, A., {Severgnini}, P., {Della Ceca}, R., {Maccacaro}, T.,
  {Carrera}, F.~J., \& {Page}, M.~J. 2007, \aap, 470, 557

\bibitem[{{Chen} {et~al.}(2013){Chen}, {Hickox}, {Alberts}, {Brodwin}, {Jones},
  {Murray}, {Alexander}, {Assef}, {Brown}, {Dey}, {Forman}, {Gorjian},
  {Goulding}, {Le Floc'h}, {Jannuzi}, {Mullaney}, \& {Pope}}]{Chen2013}
{Chen}, C.-T.~J., {Hickox}, R.~C., {Alberts}, S., {Brodwin}, M., {Jones}, C.,
  {Murray}, S.~S., {Alexander}, D.~M., {Assef}, R.~J., {Brown}, M.~J.~I.,
  {Dey}, A., {Forman}, W.~R., {Gorjian}, V., {Goulding}, A.~D., {Le Floc'h},
  E., {Jannuzi}, B.~T., {Mullaney}, J.~R., \& {Pope}, A. 2013, \apj, 773, 3

\bibitem[{{Coppi}(2003)}]{Coppi2003}
{Coppi}, P. 2003, in American Institute of Physics Conference Series, Vol. 686,
  The Astrophysics of Gravitational Wave Sources, ed. J.~M. {Centrella},
  141--150

\bibitem[{{Dalla Bont{\`a}} {et~al.}(2009){Dalla Bont{\`a}}, {Ferrarese},
  {Corsini}, {Miralda-Escud{\'e}}, {Coccato}, {Sarzi}, {Pizzella}, \&
  {Beifiori}}]{DallaBonta2009}
{Dalla Bont{\`a}}, E., {Ferrarese}, L., {Corsini}, E.~M., {Miralda-Escud{\'e}},
  J., {Coccato}, L., {Sarzi}, M., {Pizzella}, A., \& {Beifiori}, A. 2009, \apj,
  690, 537

\bibitem[{{de Vaucouleurs} {et~al.}(1991){de Vaucouleurs}, {de Vaucouleurs},
  {Corwin}, {Buta}, {Paturel}, \& {Fouqu{\'e}}}]{deVaucouleurs1991}
{de Vaucouleurs}, G., {de Vaucouleurs}, A., {Corwin}, Jr., H.~G., {Buta},
  R.~J., {Paturel}, G., \& {Fouqu{\'e}}, P. 1991, {Third Reference Catalogue of
  Bright Galaxies. Volume I: Explanations and references. Volume II: Data for
  galaxies between 0$^{h}$ and 12$^{h}$. Volume III: Data for galaxies between
  12$^{h}$ and 24$^{h}$.}

\bibitem[{{Dong} \& {De Robertis}(2006)}]{Dong2006}
{Dong}, X.~Y., \& {De Robertis}, M.~M. 2006, \aj, 131, 1236

\bibitem[{{Erwin} {et~al.}(2003){Erwin}, {Beltr{\'a}n}, {Graham}, \&
  {Beckman}}]{Erwin2003}
{Erwin}, P., {Beltr{\'a}n}, J.~C.~V., {Graham}, A.~W., \& {Beckman}, J.~E.
  2003, \apj, 597, 929

\bibitem[{{Erwin} {et~al.}(2004){Erwin}, {Graham}, \& {Caon}}]{Erwin2004}
{Erwin}, P., {Graham}, A.~W., \& {Caon}, N. 2004, Coevolution of Black Holes
  and Galaxies

\bibitem[{{Faber} {et~al.}(1997){Faber}, {Tremaine}, {Ajhar}, {Byun},
  {Dressler}, {Gebhardt}, {Grillmair}, {Kormendy}, {Lauer}, \&
  {Richstone}}]{Faber1997}
{Faber}, S.~M., {Tremaine}, S., {Ajhar}, E.~A., {Byun}, Y.-I., {Dressler}, A.,
  {Gebhardt}, K., {Grillmair}, C., {Kormendy}, J., {Lauer}, T.~R., \&
  {Richstone}, D. 1997, \aj, 114, 1771

\bibitem[{{Ferrarese} \& {Merritt}(2000)}]{Ferrarese2000}
{Ferrarese}, L., \& {Merritt}, D. 2000, \apjl, 539, L9

\bibitem[{{Fukazawa} {et~al.}(2011){Fukazawa}, {Hiragi}, {Mizuno}, {Nishino},
  {Hayashi}, {Yamasaki}, {Shirai}, {Takahashi}, \& {Ohno}}]{Fukazawa2011}
{Fukazawa}, Y., {Hiragi}, K., {Mizuno}, M., {Nishino}, S., {Hayashi}, K.,
  {Yamasaki}, T., {Shirai}, H., {Takahashi}, H., \& {Ohno}, M. 2011, \apj, 727,
  19

\bibitem[{{Gallego} {et~al.}(1996){Gallego}, {Zamorano}, {Rego}, {Alonso}, \&
  {Vitores}}]{Gallego1996}
{Gallego}, J., {Zamorano}, J., {Rego}, M., {Alonso}, O., \& {Vitores}, A.~G.
  1996, \aaps, 120, 323

\bibitem[{{Gebhardt} {et~al.}(2000){Gebhardt}, {Bender}, {Bower}, {Dressler},
  {Faber}, {Filippenko}, {Green}, {Grillmair}, {Ho}, {Kormendy}, {Lauer},
  {Magorrian}, {Pinkney}, {Richstone}, \& {Tremaine}}]{Gebhardt2000}
{Gebhardt}, K., {Bender}, R., {Bower}, G., {Dressler}, A., {Faber}, S.~M.,
  {Filippenko}, A.~V., {Green}, R., {Grillmair}, C., {Ho}, L.~C., {Kormendy},
  J., {Lauer}, T.~R., {Magorrian}, J., {Pinkney}, J., {Richstone}, D., \&
  {Tremaine}, S. 2000, \apjl, 539, L13

\bibitem[{{Georgakakis} {et~al.}(2001){Georgakakis}, {Hopkins}, {Caulton},
  {Wiklind}, {Terlevich}, \& {Forbes}}]{Georgakakis2001}
{Georgakakis}, A., {Hopkins}, A.~M., {Caulton}, A., {Wiklind}, T., {Terlevich},
  A.~I., \& {Forbes}, D.~A. 2001, \mnras, 326, 1431

\bibitem[{{Gil de Paz} {et~al.}(2007){Gil de Paz}, {Boissier}, {Madore},
  {Seibert}, {Joe}, {Boselli}, {Wyder}, {Thilker}, {Bianchi}, {Rey}, {Rich},
  {Barlow}, {Conrow}, {Forster}, {Friedman}, {Martin}, {Morrissey}, {Neff},
  {Schiminovich}, {Small}, {Donas}, {Heckman}, {Lee}, {Milliard}, {Szalay}, \&
  {Yi}}]{GildePaz2006}
{Gil de Paz}, A., {Boissier}, S., {Madore}, B.~F., {Seibert}, M., {Joe}, Y.~H.,
  {Boselli}, A., {Wyder}, T.~K., {Thilker}, D., {Bianchi}, L., {Rey}, S.-C.,
  {Rich}, R.~M., {Barlow}, T.~A., {Conrow}, T., {Forster}, K., {Friedman},
  P.~G., {Martin}, D.~C., {Morrissey}, P., {Neff}, S.~G., {Schiminovich}, D.,
  {Small}, T., {Donas}, J., {Heckman}, T.~M., {Lee}, Y.-W., {Milliard}, B.,
  {Szalay}, A.~S., \& {Yi}, S. 2007, \apjs, 173, 185

\bibitem[{{Graham} {et~al.}(2011){Graham}, {Onken}, {Athanassoula}, \&
  {Combes}}]{Graham2011}
{Graham}, A.~W., {Onken}, C.~A., {Athanassoula}, E., \& {Combes}, F. 2011,
  \mnras, 412, 2211

\bibitem[{{Graham} \& {Scott}(2013)}]{Graham2013}
{Graham}, A.~W., \& {Scott}, N. 2013, \apj, 764, 151

\bibitem[{{Greene} {et~al.}(2010){Greene}, {Peng}, {Kim}, {Kuo}, {Braatz},
  {Impellizzeri}, {Condon}, {Lo}, {Henkel}, \& {Reid}}]{Greene2010}
{Greene}, J.~E., {Peng}, C.~Y., {Kim}, M., {Kuo}, C.-Y., {Braatz}, J.~A.,
  {Impellizzeri}, C.~M.~V., {Condon}, J.~J., {Lo}, K.~Y., {Henkel}, C., \&
  {Reid}, M.~J. 2010, \apj, 721, 26

\bibitem[{{Gu} {et~al.}(2006){Gu}, {Melnick}, {Cid Fernandes}, {Kunth},
  {Terlevich}, \& {Terlevich}}]{Gu2006}
{Gu}, Q., {Melnick}, J., {Cid Fernandes}, R., {Kunth}, D., {Terlevich}, E., \&
  {Terlevich}, R. 2006, VizieR Online Data Catalog, 736, 60480

\bibitem[{{G{\"u}ltekin} {et~al.}(2009{\natexlab{a}}){G{\"u}ltekin}, {Cackett},
  {Miller}, {Di Matteo}, {Markoff}, \& {Richstone}}]{GultekinNovember2009}
{G{\"u}ltekin}, K., {Cackett}, E.~M., {Miller}, J.~M., {Di Matteo}, T.,
  {Markoff}, S., \& {Richstone}, D.~O. 2009{\natexlab{a}}, \apj, 706, 404

\bibitem[{{G{\"u}ltekin} {et~al.}(2012){G{\"u}ltekin}, {Cackett}, {Miller}, {Di
  Matteo}, {Markoff}, \& {Richstone}}]{Gultekin2012}
---. 2012, \apj, 749, 129

\bibitem[{{G{\"u}ltekin} {et~al.}(2009{\natexlab{b}}){G{\"u}ltekin},
  {Richstone}, {Gebhardt}, {Lauer}, {Tremaine}, {Aller}, {Bender}, {Dressler},
  {Faber}, {Filippenko}, {Green}, {Ho}, {Kormendy}, {Magorrian}, {Pinkney}, \&
  {Siopis}}]{GultekinJune2009}
{G{\"u}ltekin}, K., {Richstone}, D.~O., {Gebhardt}, K., {Lauer}, T.~R.,
  {Tremaine}, S., {Aller}, M.~C., {Bender}, R., {Dressler}, A., {Faber}, S.~M.,
  {Filippenko}, A.~V., {Green}, R., {Ho}, L.~C., {Kormendy}, J., {Magorrian},
  J., {Pinkney}, J., \& {Siopis}, C. 2009{\natexlab{b}}, \apj, 698, 198

\bibitem[{{Guti{\'e}rrez} {et~al.}(2004){Guti{\'e}rrez}, {Trujillo}, {Aguerri},
  {Graham}, \& {Caon}}]{Gutierrez2004}
{Guti{\'e}rrez}, C.~M., {Trujillo}, I., {Aguerri}, J.~A.~L., {Graham}, A.~W.,
  \& {Caon}, N. 2004, \apj, 602, 664

\bibitem[{{Haehnelt} \& {Kauffmann}(2000)}]{Haehnelt2000}
{Haehnelt}, M.~G., \& {Kauffmann}, G. 2000, \mnras, 318, L35

\bibitem[{{Hainline} {et~al.}(2012){Hainline}, {Shapley}, {Greene}, {Steidel},
  {Reddy}, \& {Erb}}]{Hainline2012}
{Hainline}, K.~N., {Shapley}, A.~E., {Greene}, J.~E., {Steidel}, C.~C.,
  {Reddy}, N.~A., \& {Erb}, D.~K. 2012, \apj, 760, 74

\bibitem[{{H{\"a}ring} \& {Rix}(2004)}]{Haring2004}
{H{\"a}ring}, N., \& {Rix}, H.-W. 2004, \apjl, 604, L89

\bibitem[{{Hickox} {et~al.}(2014){Hickox}, {Mullaney}, {Alexander}, {Chen},
  {Civano}, {Goulding}, \& {Hainline}}]{Hickox2014}
{Hickox}, R.~C., {Mullaney}, J.~R., {Alexander}, D.~M., {Chen}, C.-T.~J.,
  {Civano}, F.~M., {Goulding}, A.~D., \& {Hainline}, K.~N. 2014, \apj, 782, 9

\bibitem[{{Ho} {et~al.}(2003){Ho}, {Filippenko}, \& {Sargent}}]{Ho2003}
{Ho}, L.~C., {Filippenko}, A.~V., \& {Sargent}, W.~L.~W. 2003, \apj, 583, 159

\bibitem[{{Hu}(2008)}]{Hu2008}
{Hu}, J. 2008, \mnras, 386, 2242

\bibitem[{{Hu}(2009)}]{Hu2009}
---. 2009, ArXiv e-prints

\bibitem[{{Jahnke} {et~al.}(2009){Jahnke}, {Bongiorno}, {Brusa}, {Capak},
  {Cappelluti}, {Cisternas}, {Civano}, {Colbert}, {Comastri}, {Elvis},
  {Hasinger}, {Ilbert}, {Impey}, {Inskip}, {Koekemoer}, {Lilly}, {Maier},
  {Merloni}, {Riechers}, {Salvato}, {Schinnerer}, {Scoville}, {Silverman},
  {Taniguchi}, {Trump}, \& {Yan}}]{Jahnke2009}
{Jahnke}, K., {Bongiorno}, A., {Brusa}, M., {Capak}, P., {Cappelluti}, N.,
  {Cisternas}, M., {Civano}, F., {Colbert}, J., {Comastri}, A., {Elvis}, M.,
  {Hasinger}, G., {Ilbert}, O., {Impey}, C., {Inskip}, K., {Koekemoer}, A.~M.,
  {Lilly}, S., {Maier}, C., {Merloni}, A., {Riechers}, D., {Salvato}, M.,
  {Schinnerer}, E., {Scoville}, N.~Z., {Silverman}, J., {Taniguchi}, Y.,
  {Trump}, J.~R., \& {Yan}, L. 2009, \apjl, 706, L215

\bibitem[{{Kailey} \& {Lebofsky}(1988)}]{Kailey1988}
{Kailey}, W.~F., \& {Lebofsky}, M.~J. 1988, \apj, 326, 653

\bibitem[{{Kennicutt}(1998)}]{Kennicutt1998}
{Kennicutt}, Jr., R.~C. 1998, \araa, 36, 189

\bibitem[{{Khorunzhev} {et~al.}(2012){Khorunzhev}, {Sazonov}, {Burenin}, \&
  {Tkachenko}}]{Khorunzhev2012}
{Khorunzhev}, G.~A., {Sazonov}, S.~Y., {Burenin}, R.~A., \& {Tkachenko}, A.~Y.
  2012, Astronomy Letters, 38, 475

\bibitem[{{Kormendy} {et~al.}(2011){Kormendy}, {Bender}, \&
  {Cornell}}]{Kormendy2011}
{Kormendy}, J., {Bender}, R., \& {Cornell}, M.~E. 2011, \nat, 469, 374

\bibitem[{{Kormendy} {et~al.}(2010){Kormendy}, {Drory}, {Bender}, \&
  {Cornell}}]{Kormendy2010}
{Kormendy}, J., {Drory}, N., {Bender}, R., \& {Cornell}, M.~E. 2010, \apj, 723,
  54

\bibitem[{{Kormendy} \& {Gebhardt}(2001)}]{Kormendy2001}
{Kormendy}, J., \& {Gebhardt}, K. 2001, in American Institute of Physics
  Conference Series, Vol. 586, 20th Texas Symposium on relativistic
  astrophysics, ed. J.~C. {Wheeler} \& H.~{Martel}, 363--381

\bibitem[{{Kormendy} \& {Ho}(2013)}]{KormendyandHo2013}
{Kormendy}, J., \& {Ho}, L.~C. 2013, ArXiv e-prints

\bibitem[{{Kormendy} \& {Richstone}(1995)}]{Kormendy1995}
{Kormendy}, J., \& {Richstone}, D. 1995, \araa, 33, 581

\bibitem[{{Kuo} {et~al.}(2011){Kuo}, {Braatz}, {Condon}, {Impellizzeri}, {Lo},
  {Zaw}, {Schenker}, {Henkel}, {Reid}, \& {Greene}}]{Kuo2011}
{Kuo}, C.~Y., {Braatz}, J.~A., {Condon}, J.~J., {Impellizzeri}, C.~M.~V., {Lo},
  K.~Y., {Zaw}, I., {Schenker}, M., {Henkel}, C., {Reid}, M.~J., \& {Greene},
  J.~E. 2011, \apj, 727, 20

\bibitem[{{Lauer} {et~al.}(2007){Lauer}, {Tremaine}, {Richstone}, \&
  {Faber}}]{Lauer2007}
{Lauer}, T.~R., {Tremaine}, S., {Richstone}, D., \& {Faber}, S.~M. 2007, \apj,
  670, 249

\bibitem[{{Lusso} {et~al.}(2011){Lusso}, {Comastri}, {Vignali}, {Zamorani},
  {Treister}, {Sanders}, {Bolzonella}, {Bongiorno}, {Brusa}, {Civano}, {Gilli},
  {Mainieri}, {Nair}, {Aller}, {Carollo}, {Koekemoer}, {Merloni}, \&
  {Trump}}]{Lusso2011}
{Lusso}, E., {Comastri}, A., {Vignali}, C., {Zamorani}, G., {Treister}, E.,
  {Sanders}, D., {Bolzonella}, M., {Bongiorno}, A., {Brusa}, M., {Civano}, F.,
  {Gilli}, R., {Mainieri}, V., {Nair}, P., {Aller}, M.~C., {Carollo}, M.,
  {Koekemoer}, A.~M., {Merloni}, A., \& {Trump}, J.~R. 2011, \aap, 534, A110

\bibitem[{{Magorrian} {et~al.}(1998){Magorrian}, {Tremaine}, {Richstone},
  {Bender}, {Bower}, {Dressler}, {Faber}, {Gebhardt}, {Green}, {Grillmair},
  {Kormendy}, \& {Lauer}}]{Magorrian1998}
{Magorrian}, J., {Tremaine}, S., {Richstone}, D., {Bender}, R., {Bower}, G.,
  {Dressler}, A., {Faber}, S.~M., {Gebhardt}, K., {Green}, R., {Grillmair}, C.,
  {Kormendy}, J., \& {Lauer}, T. 1998, \aj, 115, 2285

\bibitem[{{Marconi} \& {Hunt}(2003)}]{Marconi2003}
{Marconi}, A., \& {Hunt}, L.~K. 2003, \apjl, 589, L21

\bibitem[{{Mazzuca} {et~al.}(2008){Mazzuca}, {Knapen}, {Veilleux}, \&
  {Regan}}]{Mazzuca2008}
{Mazzuca}, L.~M., {Knapen}, J.~H., {Veilleux}, S., \& {Regan}, M.~W. 2008,
  \apjs, 174, 337

\bibitem[{{McConnell} \& {Ma}(2013)}]{McConnellMa2013}
{McConnell}, N.~J., \& {Ma}, C.-P. 2013, \apj, 764, 184

\bibitem[{{McLure} \& {Dunlop}(2002)}]{McLureDunlop2002}
{McLure}, R.~J., \& {Dunlop}, J.~S. 2002, \mnras, 331, 795

\bibitem[{{Mel{\'e}ndez} {et~al.}(2008){Mel{\'e}ndez}, {Kraemer}, {Schmitt},
  {Crenshaw}, {Deo}, {Mushotzky}, \& {Bruhweiler}}]{Melendez2008}
{Mel{\'e}ndez}, M., {Kraemer}, S.~B., {Schmitt}, H.~R., {Crenshaw}, D.~M.,
  {Deo}, R.~P., {Mushotzky}, R.~F., \& {Bruhweiler}, F.~C. 2008, \apj, 689, 95

\bibitem[{{Merloni} {et~al.}(2010){Merloni}, {Bongiorno}, {Bolzonella},
  {Brusa}, {Civano}, {Comastri}, {Elvis}, {Fiore}, {Gilli}, {Hao}, {Jahnke},
  {Koekemoer}, {Lusso}, {Mainieri}, {Mignoli}, {Miyaji}, {Renzini}, {Salvato},
  {Silverman}, {Trump}, {Vignali}, {Zamorani}, {Capak}, {Lilly}, {Sanders},
  {Taniguchi}, {Bardelli}, {Carollo}, {Caputi}, {Contini}, {Coppa}, {Cucciati},
  {de la Torre}, {de Ravel}, {Franzetti}, {Garilli}, {Hasinger}, {Impey},
  {Iovino}, {Iwasawa}, {Kampczyk}, {Kneib}, {Knobel}, {Kova{\v c}},
  {Lamareille}, {Le Borgne}, {Le Brun}, {Le F{\`e}vre}, {Maier}, {Pello},
  {Peng}, {Perez Montero}, {Ricciardelli}, {Scodeggio}, {Tanaka}, {Tasca},
  {Tresse}, {Vergani}, \& {Zucca}}]{Merloni2010}
{Merloni}, A., {Bongiorno}, A., {Bolzonella}, M., {Brusa}, M., {Civano}, F.,
  {Comastri}, A., {Elvis}, M., {Fiore}, F., {Gilli}, R., {Hao}, H., {Jahnke},
  K., {Koekemoer}, A.~M., {Lusso}, E., {Mainieri}, V., {Mignoli}, M., {Miyaji},
  T., {Renzini}, A., {Salvato}, M., {Silverman}, J., {Trump}, J., {Vignali},
  C., {Zamorani}, G., {Capak}, P., {Lilly}, S.~J., {Sanders}, D., {Taniguchi},
  Y., {Bardelli}, S., {Carollo}, C.~M., {Caputi}, K., {Contini}, T., {Coppa},
  G., {Cucciati}, O., {de la Torre}, S., {de Ravel}, L., {Franzetti}, P.,
  {Garilli}, B., {Hasinger}, G., {Impey}, C., {Iovino}, A., {Iwasawa}, K.,
  {Kampczyk}, P., {Kneib}, J.-P., {Knobel}, C., {Kova{\v c}}, K., {Lamareille},
  F., {Le Borgne}, J.-F., {Le Brun}, V., {Le F{\`e}vre}, O., {Maier}, C.,
  {Pello}, R., {Peng}, Y., {Perez Montero}, E., {Ricciardelli}, E.,
  {Scodeggio}, M., {Tanaka}, M., {Tasca}, L.~A.~M., {Tresse}, L., {Vergani},
  D., \& {Zucca}, E. 2010, \apj, 708, 137

\bibitem[{{Mullaney} {et~al.}(2012{\natexlab{a}}){Mullaney}, {Daddi},
  {B{\'e}thermin}, {Elbaz}, {Juneau}, {Pannella}, {Sargent}, {Alexander}, \&
  {Hickox}}]{Mullaney2012b}
{Mullaney}, J.~R., {Daddi}, E., {B{\'e}thermin}, M., {Elbaz}, D., {Juneau}, S.,
  {Pannella}, M., {Sargent}, M.~T., {Alexander}, D.~M., \& {Hickox}, R.~C.
  2012{\natexlab{a}}, \apjl, 753, L30

\bibitem[{{Mullaney} {et~al.}(2012{\natexlab{b}}){Mullaney}, {Daddi},
  {B{\'e}thermin}, {Elbaz}, {Juneau}, {Pannella}, {Sargent}, {Alexander}, \&
  {Hickox}}]{Mullaney2012}
---. 2012{\natexlab{b}}, \apjl, 753, L30

\bibitem[{{Mullaney} {et~al.}(2012{\natexlab{c}}){Mullaney}, {Pannella},
  {Daddi}, {Alexander}, {Elbaz}, {Hickox}, {Bournaud}, {Altieri}, {Aussel},
  {Coia}, {Dannerbauer}, {Dasyra}, {Dickinson}, {Hwang}, {Kartaltepe},
  {Leiton}, {Magdis}, {Magnelli}, {Popesso}, {Valtchanov}, {Bauer}, {Brandt},
  {Del Moro}, {Hanish}, {Ivison}, {Juneau}, {Luo}, {Lutz}, {Sargent}, {Scott},
  \& {Xue}}]{Mullaney2012a}
{Mullaney}, J.~R., {Pannella}, M., {Daddi}, E., {Alexander}, D.~M., {Elbaz},
  D., {Hickox}, R.~C., {Bournaud}, F., {Altieri}, B., {Aussel}, H., {Coia}, D.,
  {Dannerbauer}, H., {Dasyra}, K., {Dickinson}, M., {Hwang}, H.~S.,
  {Kartaltepe}, J., {Leiton}, R., {Magdis}, G., {Magnelli}, B., {Popesso}, P.,
  {Valtchanov}, I., {Bauer}, F.~E., {Brandt}, W.~N., {Del Moro}, A., {Hanish},
  D.~J., {Ivison}, R.~J., {Juneau}, S., {Luo}, B., {Lutz}, D., {Sargent},
  M.~T., {Scott}, D., \& {Xue}, Y.~Q. 2012{\natexlab{c}}, \mnras, 419, 95

\bibitem[{{Novak} {et~al.}(2006){Novak}, {Faber}, \& {Dekel}}]{Novak2006}
{Novak}, G.~S., {Faber}, S.~M., \& {Dekel}, A. 2006, \apj, 637, 96

\bibitem[{{Peng} {et~al.}(2006){Peng}, {Impey}, {Rix}, {Kochanek}, {Keeton},
  {Falco}, {Leh{\'a}r}, \& {McLeod}}]{Peng2006}
{Peng}, C.~Y., {Impey}, C.~D., {Rix}, H.-W., {Kochanek}, C.~S., {Keeton},
  C.~R., {Falco}, E.~E., {Leh{\'a}r}, J., \& {McLeod}, B.~A. 2006, \apj, 649,
  616

\bibitem[{{P{\'e}rez-Gonz{\'a}lez} {et~al.}(2000){P{\'e}rez-Gonz{\'a}lez},
  {Zamorano}, {Gallego}, \& {Gil de Paz}}]{PerezGonzales2000}
{P{\'e}rez-Gonz{\'a}lez}, P.~G., {Zamorano}, J., {Gallego}, J., \& {Gil de
  Paz}, A. 2000, \aaps, 141, 409

\bibitem[{{Prugniel} {et~al.}(2011){Prugniel}, {Zeilinger}, {Koleva}, \& {de
  Rijcke}}]{Prugniel2011}
{Prugniel}, P., {Zeilinger}, W., {Koleva}, M., \& {de Rijcke}, S. 2011, \aap,
  528, A128

\bibitem[{{Robitaille} \& {Whitney}(2010)}]{Robitaille2010}
{Robitaille}, T.~P., \& {Whitney}, B.~A. 2010, \apjl, 710, L11

\bibitem[{{Rosario} {et~al.}(2013{\natexlab{a}}){Rosario}, {Mozena}, {Wuyts},
  {Nandra}, {Koekemoer}, {McGrath}, {Hathi}, {Dekel}, {Donley}, {Dunlop},
  {Faber}, {Ferguson}, {Giavalisco}, {Grogin}, {Guo}, {Kocevski}, {Koo},
  {Laird}, {Newman}, {Rangel}, \& {Somerville}}]{Rosario2013a}
{Rosario}, D.~J., {Mozena}, M., {Wuyts}, S., {Nandra}, K., {Koekemoer}, A.,
  {McGrath}, E., {Hathi}, N.~P., {Dekel}, A., {Donley}, J., {Dunlop}, J.~S.,
  {Faber}, S.~M., {Ferguson}, H., {Giavalisco}, M., {Grogin}, N., {Guo}, Y.,
  {Kocevski}, D.~D., {Koo}, D.~C., {Laird}, E., {Newman}, J., {Rangel}, C., \&
  {Somerville}, R. 2013{\natexlab{a}}, \apj, 763, 59

\bibitem[{{Rosario} {et~al.}(2013{\natexlab{b}}){Rosario}, {Santini}, {Lutz},
  {Netzer}, {Bauer}, {Berta}, {Magnelli}, {Popesso}, {Alexander}, {Brandt},
  {Genzel}, {Maiolino}, {Mullaney}, {Nordon}, {Saintonge}, {Tacconi}, \&
  {Wuyts}}]{Rosario2013b}
{Rosario}, D.~J., {Santini}, P., {Lutz}, D., {Netzer}, H., {Bauer}, F.~E.,
  {Berta}, S., {Magnelli}, B., {Popesso}, P., {Alexander}, D.~M., {Brandt},
  W.~N., {Genzel}, R., {Maiolino}, R., {Mullaney}, J.~R., {Nordon}, R.,
  {Saintonge}, A., {Tacconi}, L., \& {Wuyts}, S. 2013{\natexlab{b}}, \apj, 771,
  63

\bibitem[{{Sani} {et~al.}(2011){Sani}, {Marconi}, {Hunt}, \&
  {Risaliti}}]{Sani2011}
{Sani}, E., {Marconi}, A., {Hunt}, L.~K., \& {Risaliti}, G. 2011, \mnras, 413,
  1479

\bibitem[{{Sargsyan} {et~al.}(2011){Sargsyan}, {Weedman}, {Lebouteiller},
  {Houck}, {Barry}, {Hovhannisyan}, \& {Mickaelian}}]{Sargsyan2011}
{Sargsyan}, L., {Weedman}, D., {Lebouteiller}, V., {Houck}, J., {Barry}, D.,
  {Hovhannisyan}, A., \& {Mickaelian}, A. 2011, \apj, 730, 19

\bibitem[{{Schmitt} {et~al.}(2003){Schmitt}, {Donley}, {Antonucci},
  {Hutchings}, \& {Kinney}}]{Schmitt2003}
{Schmitt}, H.~R., {Donley}, J.~L., {Antonucci}, R.~R.~J., {Hutchings}, J.~B.,
  \& {Kinney}, A.~L. 2003, \apjs, 148, 327

\bibitem[{{Schramm} \& {Silverman}(2013)}]{SchrammSilverman2013}
{Schramm}, M., \& {Silverman}, J.~D. 2013, \apj, 767, 13

\bibitem[{{Shapley} {et~al.}(2001){Shapley}, {Fabbiano}, \&
  {Eskridge}}]{Shapley2001}
{Shapley}, A., {Fabbiano}, G., \& {Eskridge}, P.~B. 2001, \apjs, 137, 139

\bibitem[{{Shen} \& {Kelly}(2010)}]{ShenKelly2010}
{Shen}, Y., \& {Kelly}, B.~C. 2010, \apj, 713, 41

\bibitem[{{Silverman} {et~al.}(2009){Silverman}, {Lamareille}, {Maier},
  {Lilly}, {Mainieri}, {Brusa}, {Cappelluti}, {Hasinger}, {Zamorani},
  {Scodeggio}, {Bolzonella}, {Contini}, {Carollo}, {Jahnke}, {Kneib}, {Le
  F{\`e}vre}, {Merloni}, {Bardelli}, {Bongiorno}, {Brunner}, {Caputi},
  {Civano}, {Comastri}, {Coppa}, {Cucciati}, {de la Torre}, {de Ravel},
  {Elvis}, {Finoguenov}, {Fiore}, {Franzetti}, {Garilli}, {Gilli}, {Iovino},
  {Kampczyk}, {Knobel}, {Kova{\v c}}, {Le Borgne}, {Le Brun}, {Mignoli},
  {Pello}, {Peng}, {Perez Montero}, {Ricciardelli}, {Tanaka}, {Tasca},
  {Tresse}, {Vergani}, {Vignali}, {Zucca}, {Bottini}, {Cappi}, {Cassata},
  {Fumana}, {Griffiths}, {Kartaltepe}, {Koekemoer}, {Marinoni}, {McCracken},
  {Memeo}, {Meneux}, {Oesch}, {Porciani}, \& {Salvato}}]{Silverman2009}
{Silverman}, J.~D., {Lamareille}, F., {Maier}, C., {Lilly}, S.~J., {Mainieri},
  V., {Brusa}, M., {Cappelluti}, N., {Hasinger}, G., {Zamorani}, G.,
  {Scodeggio}, M., {Bolzonella}, M., {Contini}, T., {Carollo}, C.~M., {Jahnke},
  K., {Kneib}, J.-P., {Le F{\`e}vre}, O., {Merloni}, A., {Bardelli}, S.,
  {Bongiorno}, A., {Brunner}, H., {Caputi}, K., {Civano}, F., {Comastri}, A.,
  {Coppa}, G., {Cucciati}, O., {de la Torre}, S., {de Ravel}, L., {Elvis}, M.,
  {Finoguenov}, A., {Fiore}, F., {Franzetti}, P., {Garilli}, B., {Gilli}, R.,
  {Iovino}, A., {Kampczyk}, P., {Knobel}, C., {Kova{\v c}}, K., {Le Borgne},
  J.-F., {Le Brun}, V., {Mignoli}, M., {Pello}, R., {Peng}, Y., {Perez
  Montero}, E., {Ricciardelli}, E., {Tanaka}, M., {Tasca}, L., {Tresse}, L.,
  {Vergani}, D., {Vignali}, C., {Zucca}, E., {Bottini}, D., {Cappi}, A.,
  {Cassata}, P., {Fumana}, M., {Griffiths}, R., {Kartaltepe}, J., {Koekemoer},
  A., {Marinoni}, C., {McCracken}, H.~J., {Memeo}, P., {Meneux}, B., {Oesch},
  P., {Porciani}, C., \& {Salvato}, M. 2009, \apj, 696, 396

\bibitem[{{Sun} {et~al.}(2014){Sun}, {Trump}, {Brandt}, {Luo}, {Alexander},
  {Jahnke}, {Rosario}, {Wang}, \& {Xue}}]{Sun2014}
{Sun}, M., {Trump}, J.~R., {Brandt}, W.~N., {Luo}, B., {Alexander}, D.,
  {Jahnke}, K., {Rosario}, D.~J., {Wang}, S.~X., \& {Xue}, Y.~Q. 2014,
  submitted

\bibitem[{{Tremaine} {et~al.}(2002){Tremaine}, {Gebhardt}, {Bender}, {Bower},
  {Dressler}, {Faber}, {Filippenko}, {Green}, {Grillmair}, {Ho}, {Kormendy},
  {Lauer}, {Magorrian}, {Pinkney}, \& {Richstone}}]{Tremaine2002}
{Tremaine}, S., {Gebhardt}, K., {Bender}, R., {Bower}, G., {Dressler}, A.,
  {Faber}, S.~M., {Filippenko}, A.~V., {Green}, R., {Grillmair}, C., {Ho},
  L.~C., {Kormendy}, J., {Lauer}, T.~R., {Magorrian}, J., {Pinkney}, J., \&
  {Richstone}, D. 2002, \apj, 574, 740

\bibitem[{{Treu} {et~al.}(2007){Treu}, {Woo}, {Malkan}, \&
  {Blandford}}]{Treu2007}
{Treu}, T., {Woo}, J.-H., {Malkan}, M.~A., \& {Blandford}, R.~D. 2007, \apj,
  667, 117

\bibitem[{{Vika} {et~al.}(2012){Vika}, {Driver}, {Cameron}, {Kelvin}, \&
  {Robotham}}]{Vika2012}
{Vika}, M., {Driver}, S.~P., {Cameron}, E., {Kelvin}, L., \& {Robotham}, A.
  2012, \mnras, 419, 2264

\bibitem[{{Wu} {et~al.}(2009){Wu}, {Charmandaris}, {Huang}, {Spinoglio}, \&
  {Tommasin}}]{Wu2009}
{Wu}, Y., {Charmandaris}, V., {Huang}, J., {Spinoglio}, L., \& {Tommasin}, S.
  2009, \apj, 701, 658

\bibitem[{{Zhu} {et~al.}(2012){Zhu}, {Li}, \& {Sherman}}]{Zhu2012}
{Zhu}, Q., {Li}, Y., \& {Sherman}, S. 2012, ArXiv e-prints

\end{thebibliography}
\end{document}